\documentclass[amssymb,aip,sd,preprint,pdftex,colorlinks=true,linkcolor=blue,urlcolor=blue]{revtex4-1}

\usepackage{amsmath,color}
\usepackage{amssymb}
\usepackage{amsfonts}

\usepackage{graphicx}
\usepackage{framed}
\usepackage{hyperref}

\newcommand{\da}{^\dagger}

\newcommand{\dfdx}[2]{\dfrac{\partial#1}{\partial#2}}

\newcommand{\br}[1]{\langle #1 \vert}
\newcommand{\ke}[1]{\vert #1  \rangle}
\newcommand{\bk}[2]{\langle #1  \vert #2  \rangle}

\renewcommand{\vec}[1]{\mathbf{#1}}

\begin{document}

\title{Monitoring Nonadiabatic Avoided Crossing Dynamics in Molecules by Ultrafast X-Ray Diffraction}
\author{Markus Kowalewski$^{a,b}$,$^\dagger$}
\email{mkowalew@uci.edu}
\author{Kochise Bennett$^{b}$}
\thanks{These authors contributed equally to this manuscript}
\author{Shaul Mukamel$^{c}$}
\email{smukamel@uci.edu}
\affiliation{Chemistry Department, University of California, Irvine, California 92697-2025, USA}
\affiliation{Department of Physics and Astronomy, University of California, Irvine, California 92697-2025, USA}

\date{\today}
\begin{abstract}
We examine time-resolved X-ray diffraction from molecules in the gas phase
which undergo nonadiabatic avoided-crossing dynamics involving strongly coupled electrons and nuclei. Several contributions to the signal are identified, representing
(in decreasing strength) elastic scattering,
contributions of the electronic coherences created by nonadiabatic couplings in the avoided crossing regime, and inelastic scattering.
The former probes the  charge density and delivers direct information on the evolving
molecular geometry. The latter two contributions are weaker and carry spatial information of
the transition charge densities (off-diagonal elements of the
charge-density operator).
Simulations are presented for the nonadiabatic harpooning process in the excited states
of sodium fluoride.
\end{abstract}
\maketitle

\section{Introduction}
X-ray diffraction \cite{guiner, modxrayphys,velser} has been used for over a century to probe the structure of crystals and has been extended to diffuse scattering from liquids, probing nearest-neighbor distances, and serves as inspiration for the conceptually similar electron diffraction technique \cite{ben1996direct,siwick2003atomic}.
Time-resolved X-ray diffraction (TRXD) can track the structural changes that characterize phase transitions and chemical reactions and has been actively pursued to create movies of elementary
molecular events \cite{bratos2002time,siwick2003atomic,coppens2005structure,ihee2005ultrafast, wulff2006recombination, cammarata2008tracking,siders,woerner2010concerted, coppens2011molecular, neutze2012time,yang2016diffractive, falcone,glownia2016self,Bennett16prlcomment}.
Free electron lasers allow extremely bright and ultrafast X-ray pulses.
These should make it possible to push diffraction to the single-molecule limit,
\cite{stevens, mcpherson, hajdu, chapmanbeyond, starodub,Kahra12}
eliminating the need for time-consuming crystal
preparation.
In addition, their femtosecond
timescale opens up the possibility of tracking attosecond electronic dynamics while the brightness may permit even weak signals, such as inelastic scattering from transient electronic coherences, to be measured \cite{altarelli, feldhaus, mcneil, naturefsnano,bostedt2013ultra,barty2013molecular}.

\par
In this paper, we show how TRXD may be used to obtain real-time stroboscopic snapshots of nonadiabatic molecular dynamics.
Nonadiabatic processes control virtually all photochemical and photophysical
processes in molecules. For a single vibrational
coordinate, this results in avoided crossings. With two or more
vibrational degrees of freedom, conical intersections (CoIns)
become possible.
As a molecule passes through a conical intersection \cite{Domcke}
or avoided crossing,
a short-lived electronic coherence is created, which can be spectroscopically detected 
\cite{Kowalewski15prl,Kowalewski16prl} by X-rays.
Examples for a photochemical reactions that are mediated 
by a CoIn and has been studied by TRXD \cite{Minitti15prl} is the ring-opening reaction in cyclohexadiene \cite{Hofmann01,Tamura06jcp} and the cis/trans isomierization in the photoactive yellow protein \cite{Pande16sci}. Potential signatures in TRXD signals might also be useful to measure the geometric (Berry) phase \cite{Xiao10rmp}, which has so far eluded detection in molecules.

\par
We examine the elastic and inelastic contributions
to the diffraction pattern that stem from the coupled nonadiabatic electronic+nuclear
dynamics in the vicinity of an avoided crossing. 
Time-resolved scattering from photoexcited molecules in the gas phase is given by an incoherent sum of single-molecule contributions, contains elastic and inelastic terms, and may depend on electronic coherence \cite{Cao98jpca,Henriksen08jpcb}.
We calculate the TRXD by an ensemble of molecules prepared in a superposition of valence electronic and vibrational states. We identify five distinct contributions to the signal and study their relative intensity and time-resolved diffraction pattern.
Contributions from electronic coherences, which are
created in the avoided crossing region are of particular interest.
The underlying molecular quantities are the transition charge densities between electronic states.
The nonadiabatic dynamics of sodium iodide has been investigated in Ref.\ \cite{Lorenz10njp} which did not address the signatures of electronic coherences.
Dixit et al.\ \cite{Dixit12pnas} investigated diffraction of electronic
superpositions but did not include nuclear dynamics.
We examine the nonadiabatic coupled electronic+nuclear motions and the
signatures of electronic coherences in the diffraction signal of a similar molecule, sodium fluoride.

\section{The Interplay of Populations and Coherences in Single-Molecule Diffraction of Nonadiabatic Dynamics}
Our study starts with the following expression for
the off-resonant scattering signal
in the gas phase (see appendix \ref{app:sig} for derivation).
\begin{align}
S(\mathbf{q},t)&= N\int dt  \vert E_p(t)\vert^2\langle\hat{\sigma}\da(\mathbf{q},t)\hat{\sigma}(\mathbf{q},t)\rangle \label{eq:Sinctr2}
\end{align}
where $E_p(t)$ is the temporal envelope of the X-ray pulse, $\langle\dots\rangle$ stands for expectation value over the nuclear and electronic states, and $\hat{\sigma}(\mathbf{q},t)$ is the spatial Fourier transform of the charge-density operator.  
Note that Eq.\ (\ref{eq:Sinctr2}) comes with $\langle\hat{\sigma}\da(\mathbf{q},t)\hat{\sigma}(\mathbf{q},t)\rangle$ while
the classical equation for diffraction in crystals comes with 
$\vert \langle\hat{\sigma}(\mathbf{q})\rangle\vert^2$.

\begin{figure}
\includegraphics[width =.5\linewidth]{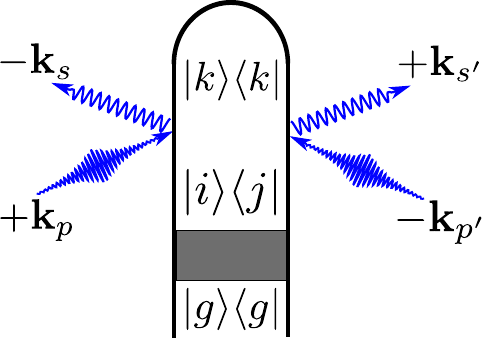}
\caption{Loop diagram for single-molecule X-ray scattering processes. The shaded area represents an excitation process that prepares the system in a superposition state by an actinic pump ($\vert g\rangle$ is the electronic ground state).  We denote modes of the X-ray pulse with $p$ and $p'$ whereas $s$, $s'$ represent relevant scattering modes ($\mathbf{k}_{p^{(\prime)}}$ has frequency $\omega_{p^{(\prime)}}$ and $\mathbf{k}_{s^{(\prime)}}$ has frequency $\omega_{s^{(\prime)}}$). Note that we use $\vert\phi_i\rangle\to\vert i\rangle$ for the electronic states in this figure to aid readability. A complete set of diagrams for Eq.\ \ref{eq:S1molExp} is given in
Fig.\ \ref{fig:Cdiag}.}
\label{fig:diagexp}
\end{figure}

\par % density
The total charge-density operator for a system composed of molecules can be written as a sum of the charge densities from each molecule
\begin{align}
\hat{\sigma}_\text{T}(\mathbf{r})=\sum_\alpha\hat{\sigma}_\alpha(\mathbf{r}-\mathbf{r}_\alpha)=\sum_\alpha\int d\mathbf{q} e^{i\mathbf{q}\cdot(\mathbf{r}-\mathbf{r}_\alpha)}\hat{\sigma}_\alpha(\mathbf{q})
\end{align}
where $\mathbf{r}_\alpha$ is the center of molecule $\alpha$. This separation is exact for a sufficiently dilute system such that the molecules have non-overlapping charge distributions, since each electron (the fundamental X-ray scatterer) can be rigorously assigned to a specific molecule.
The elastic diffraction signal from a system initially in the ground state $\vert g\rangle$ is
\begin{equation}\label{eq:SkClassic}
S(\mathbf{q})=\vert\sigma_{gg}(\mathbf{q})\vert^2,
\end{equation}
where $\sigma_{gg}(\mathbf{q})=\langle g\vert \hat{\sigma}(\mathbf{q})\vert g\rangle$ is the ground-state charge
density in $\mathbf{q}$-space and $\mathbf{q}$ is the scattering momentum transfer.
For identical molecules, the charge-density matrix elements of each
molecule only differ by the spatial phase factor associated with the location of the molecule and we may drop the subscript on $\sigma$.

We now apply these results to a molecular model consisting of two electronic states $e,g$ and a single active nuclear coordinate $R$ (Fig.\ \ref{fig:diagexp}).  The time-dependent wavefunction of each molecule in the ensemble will be expanded in the adiabatic basis
\begin{align}
\ke{\Psi(R,t)} = \sum_{i\in\{g,e\}}  c_i(t) \ke{\chi_i(R,t)} \otimes \ke{\phi_i}\end{align}
where $\ke{\chi_i(R,t)}$ is the (normalized) nuclear wave packet on the adiabatic electronic state $\ke{\phi_i}$ and $\sum_i\vert c_i\vert^2=1$ are the electronic state amplitudes.
The time evolution of $\ke{\Psi(R,t)}$ is governed by the field-free nuclear Hamiltonian $\hat{H}_0$, which includes the nonadiabatic coupling matrix elements to account for CoIns or avoided crossings in the time evolution.
The elements of the reduced electronic density matrix $\tilde\rho$ are given
by $\tilde\rho_{ij}(t) = c_i^*(t) c_j(t) \bk{\chi_i(t)}{\chi_j(t)}= \rho_{ij}\bk{\chi_i(t)}{\chi_j(t)}$, which depends
on the dephasing caused by the nuclear wave packet overlap in states $i$ and $j$.
Expanding the time-dependent densities
in the electronic states using the diagram in Fig.\ \ref{fig:diagexp} results in
\begin{align}\label{eq:S1mol}
S(\mathbf{q},t)=N \int dt \lvert E_p(t)\rvert^2 \sum_{ijk}\rho_{ij}(t)\langle \chi_i(t)\vert \hat{\sigma}\da_{ik}(\mathbf{q})\hat{\sigma}_{kj}(\mathbf{q})\vert\chi_j(t)\rangle.
\end{align} 
\begin{figure}
\includegraphics[width =1.\textwidth]{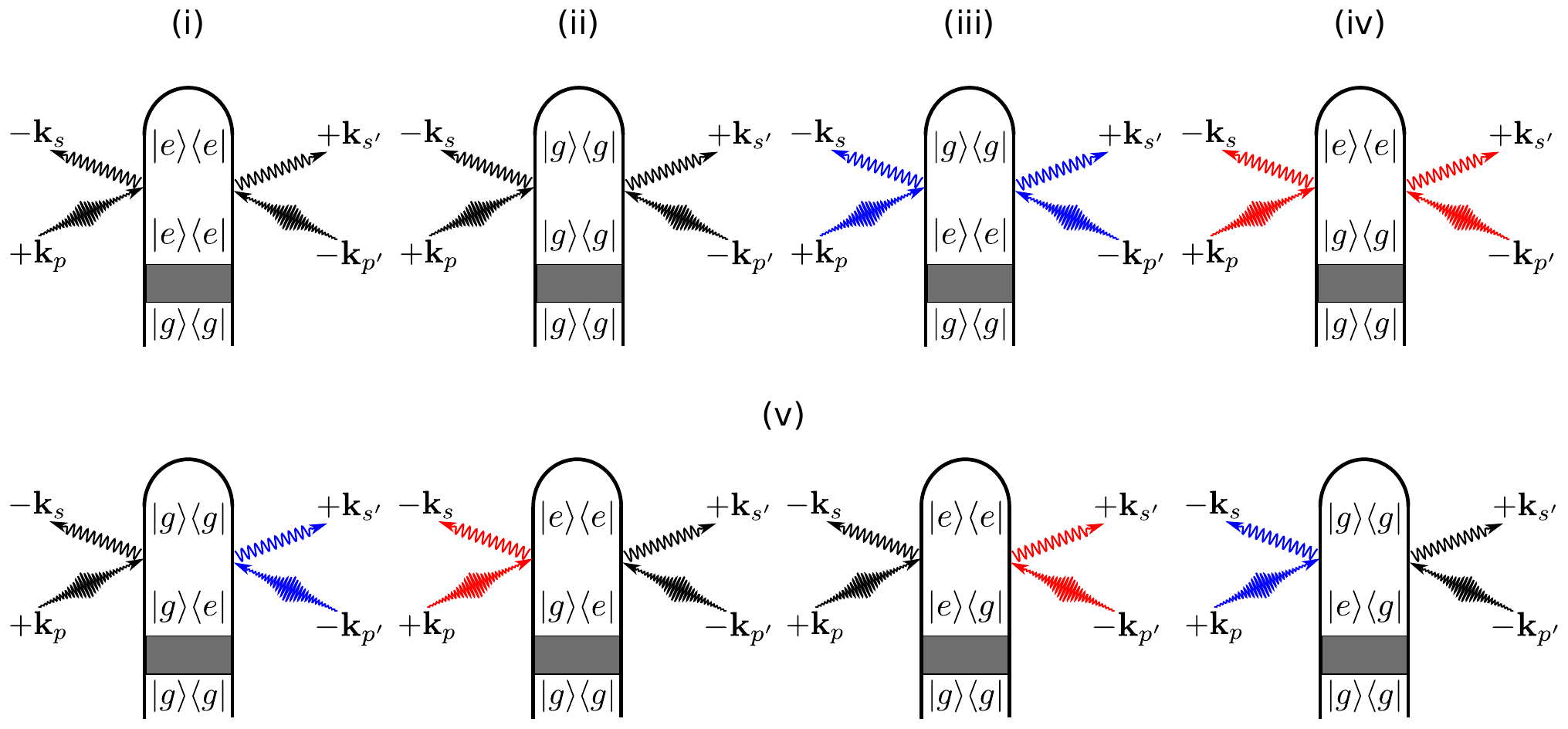}
\caption{Loop diagrams for single-molecule X-ray scattering processes as given by Eq.\ (5) and Fig. 4 in the main text.
The shaded area represents an arbitrary excitation that prepares the system in a superposition state of $\vert g\rangle$ and $\vert e\rangle$. 
Diagrams for elastic scattering from $e$ and $g$ is shown in (i) and (ii) respectively, while the diagrams for inelastic scattering from $e$ and $g$ are displayed in (iii) and (iv).
The bottom row (v) represents all diagrams involving to electronic coherences.
We denote modes of the X-ray probe pulse with $p$ and $p'$ whereas $s$, $s'$ represent relevant scattering modes ($\mathbf{k}_{p^{(\prime)}}$ has frequency $\omega_{p^{(\prime)}}$ and $\mathbf{k}_{s^{(\prime)}}$ has frequency $\omega_{s^{(\prime)}}$). Elastic scattering processes come with $\hat{\sigma}_{gg}$ or $\hat{\sigma}_{ee}$ and are denoted by black field arrows.  Inelastic processes in which the molecule gains (Stokes) or loses (anti-Stokes) energy to the field come with $\hat{\sigma}_{ge}$ or $\hat{\sigma}_{eg}$ depending whether the action is on the ket or bra and are denoted with red and blue field arrows to indicate the field's spectral shift due to the particular diagram.  Note that we use $\vert i\rangle$ instead of  $\vert\phi_i\rangle$ for the electronic states in this figure to aid readability.}
\label{fig:Cdiag}
\end{figure}
Fig. \ref{fig:Cdiag} gives the complete set of diagrams.
For a two electronic state model the diagrams in Fig. \ref{fig:Cdiag} result in the following five contributions to the signal:
\begin{widetext}
\begin{align}\label{eq:S1molExp}
\langle\hat{\sigma}\da(\mathbf{q},t)\hat{\sigma}(\mathbf{q},t)\rangle
=\bigg\lbrace
&\underbrace{\rho_{ee}(t) \br{\chi_e(t)}\hat{\sigma}\da_{ee}\hat{\sigma}_{ee}\ke{\chi_e(t)}}_{(\mathrm{i})}
+ \underbrace{\rho_{gg}(t) \br{\chi_g(t)}\hat{\sigma}\da_{gg}\hat{\sigma}_{gg}\ke{\chi_g(t)}}_{(\mathrm{ii})}  \\\nonumber
+ &\underbrace{\rho_{ee}(t) \br{\chi_e(t)}\hat{\sigma}\da_{eg}\hat{\sigma}_{ge}\ke{\chi_e(t)}}_{(\mathrm{iii})}
+ \underbrace{\rho_{gg}(t) \br{\chi_g(t)}\hat{\sigma}\da_{ge}\hat{\sigma}_{eg}\ke{\chi_g(t)}}_{(\mathrm{iv})} \\\nonumber
+&\underbrace{2\Re\big[\rho_{eg}(t)\br{\chi_e(t)}\hat{\sigma}\da_{ee}\hat{\sigma}_{eg}\ke{\chi_g(t)}
+\rho_{eg}(t)\br{\chi_e(t)}\hat{\sigma}\da_{eg}\hat{\sigma}_{gg}\ke{\chi_g(t)}\big]}_{(\mathrm{v})}\bigg\rbrace
\end{align}
\end{widetext}
where the electronic populations and coherences are given by the diagonal and off-diagonal elements of the density matrix $\rho_{ij}(t)\equiv c^*_i(t)c_j(t)$ respectively and we have defined the electronic-state matrix elements of the charge-density operator $\hat{\sigma}_{ij}\equiv\langle\phi_i\vert\hat{\sigma}(\mathbf{q})\vert\phi_j\rangle$ (which remains an operator in the nuclear space and we ommit the $\mathbf{q}$ dependence for brevity).
Equation\ (\ref{eq:S1molExp}) agrees with earlier results
\cite{Cao98jpca,Henriksen08jpcb} but identifies the  different contributions in the adiabatic basis.

\par
The first two terms on the right-hand side of Eq.\ (\ref{eq:S1molExp})
(i) and (ii) represent the elastic diffraction
from states $e$ and $g$ respectively, which encode the time evolution of the nuclear wave packets
in the two electronic states.
The next two terms, (iii) and (iv), represent the inelastic ($\sigma_{eg}^{(\dagger)}$) scattering from the electronic ground and excited state populations.
The last term (v) is due to scattering off 
electronic coherences between $\ke{g}$ and $\ke{e}$.

\par
Diffraction is often analyzed by assuming that the molecular 
electronic charge density is solely composed from the atomic densities.
In case the molecule is in the electronic state $e$ and
Eq.\ (\ref{eq:S1molExp}) can be simplified by the independent atom approximation \cite{Bartell64jacs,Waller29prs,yang2016diffractive}:
\begin{align}
\tilde S_{1,i.a.}(\vec{q})  = \sum_a \sum_{b < a} &\lvert f_a(\vec{q}) \rvert \lvert f_b(\vec{q}) \rvert
   \cos{(\phi_b(\vec{q}) - \phi_a(\vec{q}))}\nonumber\\
  &\times \int\mathrm{d}\vec{R} \mathrm{e}^{i\vec{q}\vec{R}} \chi_e^*(\vec{R}) \chi_e(\vec{R})
\end{align}
where $f_a(\vec{q})$ is the atomic charge density of the $a$th atom in the molecule and $\phi_a(\vec{q})$ is its phase factor.
This widely used expression approximates term (i) in
Eq.\ (\ref{eq:S1molExp}) but does not account 
for inelastic scattering events and contributions due to electronic coherences. Our theory explicitly seperates inelasticities, which are described by transition charge densities $\hat{\sigma}_{ij}(\mathbf{q})$ ($i\neq j$) that interfere with ground and excited state terms $\hat{\sigma}_{ii}(\mathbf{q})$.

\section{Avoided Crossing Dynamics in Sodium Fluoride}
\begin{figure}
\includegraphics[width=0.6\textwidth]{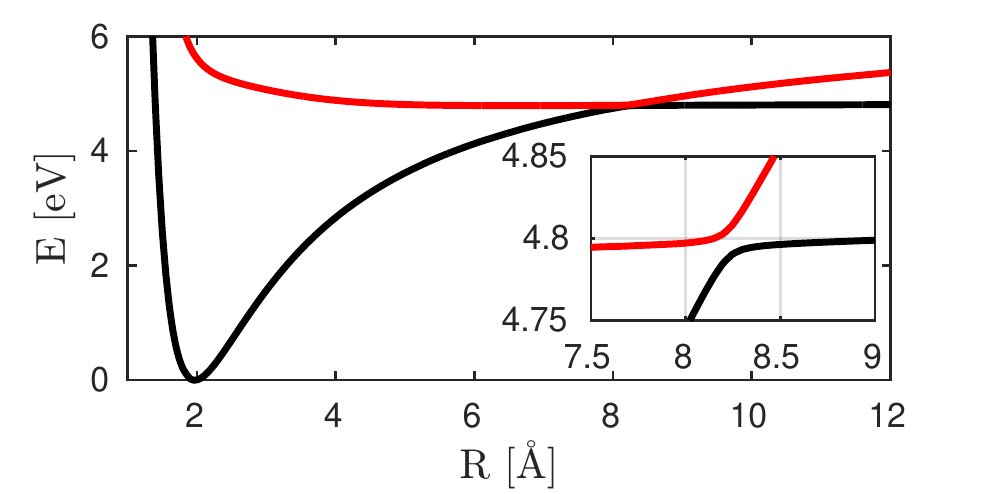}
\caption{Adiabatic potential energy surfaces for the electron harpooning in NaF (ionic $X^1\Sigma$ $\ke{g}$ black, covalent $A^1\Sigma$, $\ke{e}$ red).
The inset displays a close up of the avoided crossing region.}\label{fig:NaF_pot}
\end{figure}

We now present and discuss the five contributions to the diffraction signal from sodium fluoride.
This molecule possesses a similar electronic structure to sodium iodide, the avoided crossing of which was studied in Zewail's landmark optical experiment \cite{Rose89jcp}.
Excited-state diffraction of sodium iodide has been calculated
\cite{Lorenz10njp} by including the nonadiabatic dynamics but focusing solely on the elastic scattering processes (corresponding to terms (i) and (ii)).
An avoided crossing between the ionic and covalent state at 8\,\AA, known as  harpooning, creates an electronic coherence
in the course of the time evolution of the excited state nuclear wave packet (see Fig.\ \ref{fig:NaF_pot}).
Iodine is a strong X-ray scatterer and its large nuclear charge leads to a charge density distribution which is heavily dominated by its core electrons.
While this is still the case for molecular form factors of lighter element compounds,
they have a more prominent contribution from valence electrons compared to the core electrons. The coherence contributions, which depend on the transition densities and are dominated by the rearrangement of valence electrons are thus expected to be relatively stronger
in sodium fluoride than in sodium iodide.

\subsection{Electronic Structure Calculations and Nonadiabatic Wave packet dynamics}
The electronic structure of NaF was calculated with the program package Molpro \cite{MOLPRO_brief} at the CAS(8/9)/MRCI/aug-cc-pVTZ level of theory. 
A Douglas-Kroll-Hess 10th-order correction has been used \cite{Douglas74anp,Hess86pra} to account for relativistic effects caused by the core electrons.
All densities were evaluated from the state specific charge density matrices (and transition charge density matrices) $P^{({ij})}$, expanded in the atomic orbital basis functions $\phi_s(\vec{r})$:
\begin{align}\label{eq:sigma_qc}
\hat\sigma_{ij}(\vec{q};R) = \int \mathrm{d}\vec{r}\mathrm{e}^{-i\vec{q}\cdot\vec{r}}\sum_{rs} P^{(ij)}_{rs}(R) \phi_r^*(\vec{r};R)\phi_s(\vec{r};R)
\end{align}

\begin{figure}
\includegraphics[width=0.6\textwidth]{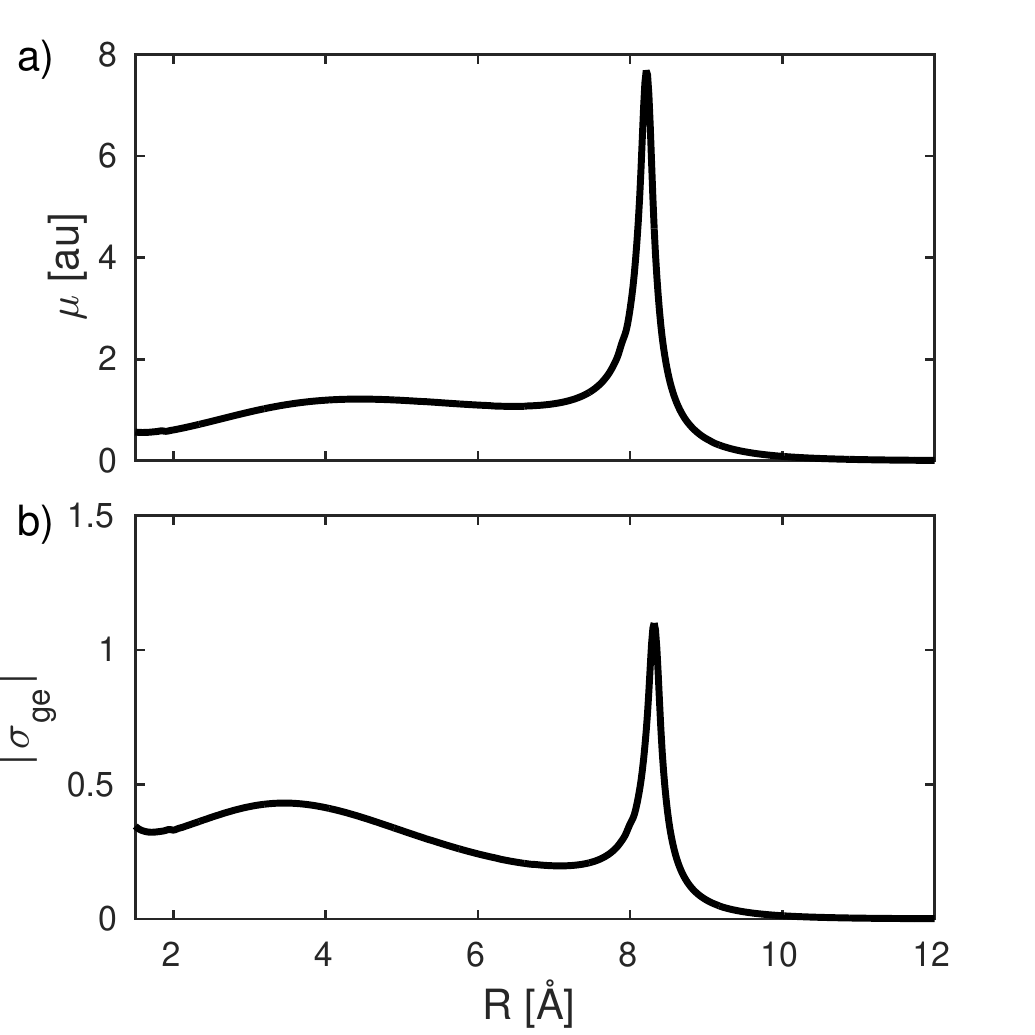}
\caption{Transition dipole moment $\mu_{ge}$ between the $X$ and $A$ states of NaF (a) and magnitude of the transition density $\sigma_{ge}$ (b).}
\label{fig:mutrans}
\end{figure}
Both the transition dipole and the integrated transition density,
$\int dr \vert \sigma_{ge} \vert$ shown in Fig. \ref{fig:mutrans}, peak at the avoided crossing point.
\begin{figure}
\includegraphics[width=0.7\textwidth]{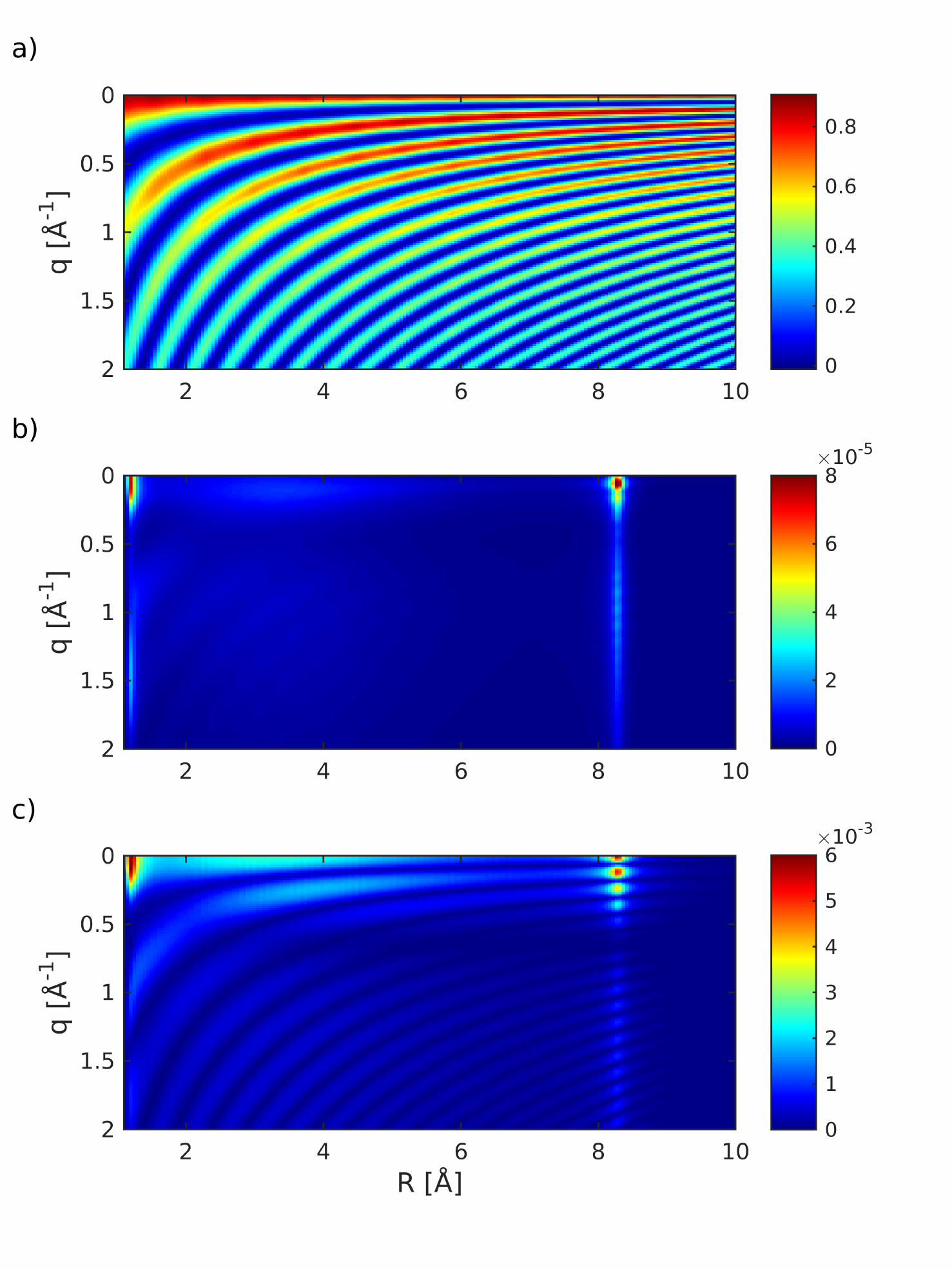}
\caption{Relevant density operator matrix elements in the nuclear subspace of NaF (obtained using Eq. (\ref{eq:sigma_qc})): (a) $\hat\sigma_{ee}^2(\vec{q},R)$, (b) $\hat\sigma_{ge}^2(\vec{q},R)$, (c) $\vert\hat\sigma_{ee}\da(\vec{q},R)\hat\sigma_{ge}(\vec{q},R)\vert$.
$\hat\sigma_{gg}^2$ is not explicitly shown due to its visual similarity to $\hat\sigma_{ee}^2$}
\label{fig:sigma_ops}
\end{figure}
The matrix elements of the electronic density operator  $\hat\sigma_{ik}^*(\vec{q};R)\hat\sigma_{kj}(\vec{q};R)$ are displayed
in Fig.\ \ref{fig:sigma_ops}. 
For clarity, only the projection along  the direction of molecular axis obtained by integrating over the perpendicular directions is shown.
The diagonal density $\hat\sigma_{ee}^2$ (Fig. \ref{fig:sigma_ops} (a)) is clearly dominated by contributions from the core electrons and the stripe pattern
reflects the bond length in reciprocal space (see Eq.\ (\ref{eq:Sinctr2})).
The transition density $\hat\sigma_{ge}^2$ (Fig.\ \ref{fig:sigma_ops} (b)) mainly contains
contributions from the valence orbitals involved in the transition. It is about 4 orders
weaker than the diagonal matrix element (Fig.\ \ref{fig:sigma_ops} (a)). However, it
peaks at the avoided crossing, making it most suitable for the detection of inelastic contributions.
The mixed matrix element $\hat\sigma_{ee}\da\hat\sigma_{eg}$ (Fig.\ \ref{fig:sigma_ops} (c)) is a product of the nuclear densities and the transition densities.

Nuclear wave packet dynamics simulations were carried out on a numerical grid
with 1200 grid points for the nuclear coordinate $R$ (extending from 2 to 24\,\AA) and the electronic states $g$ and $e$.
The Hamiltonian, which describes the
coupled electronic and vibrational degrees of freedom, is given by:
\begin{equation}
\hat H = \begin{pmatrix}
\hat T + \hat V_g(R) & - E_{pu}(t) \hat\mu_{eg}(R) + \hat K_{ge} \\
- E_{pu}(t) \hat\mu_{ge}(R) - \hat  K_{eg} & \hat T + \hat V_e(R)
\end{pmatrix}\label{eq:Hnucel}
\end{equation}
where 
\begin{align}
\hat T = -\dfrac{1}{2m} \dfdx{^2}{R^2}
\end{align}
is the kinetic operator of the nuclei, $m$ the reduced mass of the nuclei, and 
\begin{align}
\hat K_{ge} =  \dfrac{1}{2m} \left( 2f_{ge} \dfdx{}{R} +  \dfdx{}{R}  f_{ge} \right) 
\end{align}
approximates the non-adiabatic couplings \cite{Domcke,Hofmann01}.
\begin{equation}
f_{ge} = \dfrac{f_0}{(R-R_0)^2 + s^2}
\end{equation}
is the non-adiabatic coupling matrix element between $g$ and $e$ and has been obtained
by a fit to values calculated with the DDR routine in MOLPRO \cite{MOLPRO_brief}.
The fitted parameters are $f_0=0.0387$, $R_0 = 8.222$, and $s=0.0778$ (all values in atomic units).

\begin{figure}
\includegraphics[width=0.6\textwidth]{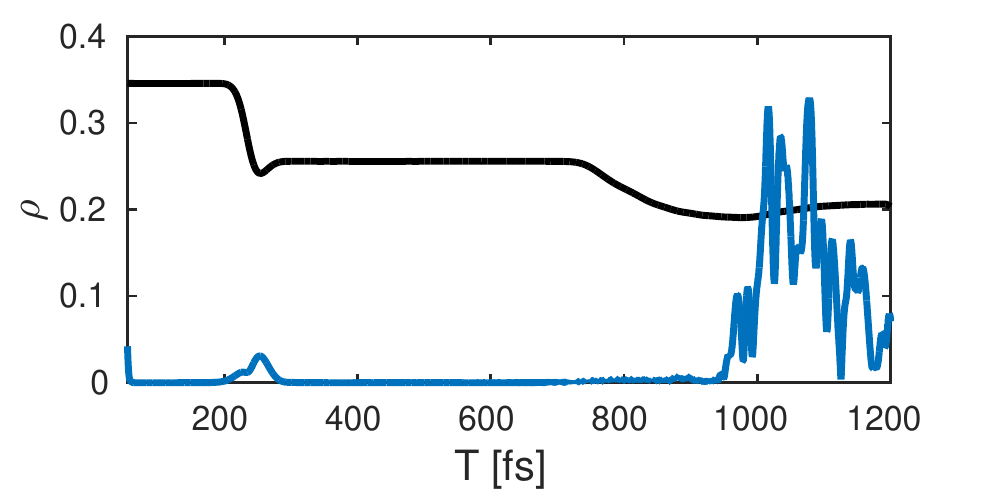}
\caption{Time evolution of the excited-state population ($A^1\Sigma$, black) and the magnitude of the coherence $\vert \rho_{eg}\vert$ (blue). The inital coherence created by the pump-pulse ($T<50$\,fs) is not shown.
The coherence at 220\,fs is created by the outward wave packet passing through the avoided crossing, while
the strong coherence around 1100\,fs corresponds to the wave packet return to
the Franck-Condon region.}\label{fig:popcoh}
\end{figure}

We assume a Gaussian pump-pulse envelope:
\begin{align}
E_{pu}(t) = E_0 \cos (\omega t) \exp(-2\ln(2)t^2   / w^2)
\end{align}
where $w$ is the full width at half maximum of the intensity profile $E_{pu}^2$.
The probe-pulse is not included in the propagation but is treated pertubatively and included in the final
signal calculation (Eq.\ (\ref{eq:Sinctr2})).
The wave function $\Psi(R,t)=(c_g(t)\chi_g(R,t),c_e(t)\chi_e(R,t))^T$ is obtained by propagating the vibrational ground state of the $X^1\Sigma$ state
with a Chebychev scheme \cite{Ezer84} using the Hamiltonian Eq.\ (\ref{eq:Hnucel}).
The kinetic operator is modified with a perfectly matched layer \cite {Nissen10} for $g$ to avoid
spurious reflections at the edges of the grid ($22$\,\AA).
The signal is then obtained by evaluating Eqs. (3) and (5) and inserting
the time-dependent wave functions and density operators ($\hat\sigma_{ik}\da\hat\sigma_{kj}$, as shown in Fig.\ \ref{fig:sigma_ops}). 
We use the adiabatic basis but the calculation is exact.
The electronic coherence is obtained from the combined
electronic-nuclear wave function as the overlap of the nuclear wave packets:
\begin{align}
\tilde\rho_{eg} = c_e^*(t) c_g(t) \bk{\chi_e}{\chi_g}
\end{align}
This results in the decay and revival of the electronic coherence.
\begin{figure}
\includegraphics[width=0.6\textwidth]{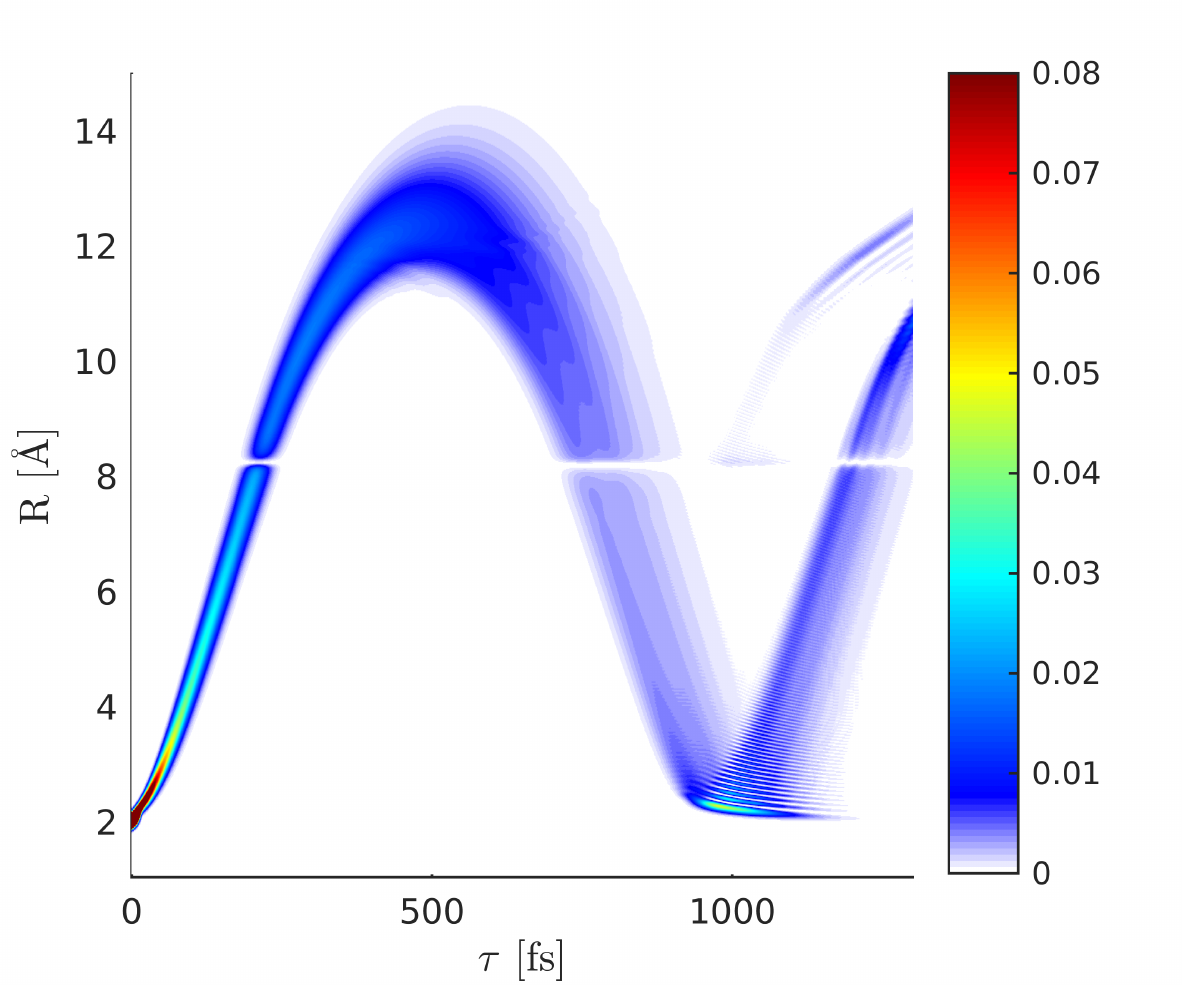}
\caption{Nuclear wave packet dynamics ($|\chi_e(R,t)|^2$) in the covalent $A^1\Sigma$ state following excitation with a 10\,fs pump-pulse (FWHM).}
\label{fig:psit2}
\end{figure}
The wave packet dynamics  in the excited state potential ($\chi_e(R,t)$) is depicted in Fig.\ \ref{fig:psit2}.
It passes through the avoided crossing between 200 and 240\,fs and reaches its outer
turning point around 500\,fs.
The time-dependent excited state population alongside with the magnitude of the electronic coherence is shown in Fig.\ \ref{fig:popcoh}. 
The two relevant valence states (Fig.\ \ref{fig:NaF_pot})
are the X$^1\Sigma^+$ ground state and the A$^1\Sigma^+$ state (referred to as
$g$ and $e$ in the following). A UV pump-pulse creates an excited-state population
$\rho_{ee}\approx 30$\,\%, triggering the nuclear wave packet dynamics in states $g$ and $e$ that is subsequently probed with a $2.5$\,fs X-ray probe pulse.
The time dependent excited-state population and
the coherence are displayed in Fig.\ \ref{fig:popcoh}. 
At around 200\,fs, the excited-state
nuclear wavepacket first reaches the avoided crossing and returns to the crossing
between 750 and 900\,fs.

\section{The Diffraction Signal}
\begin{figure}
\centering
\includegraphics[width=0.48\textwidth]{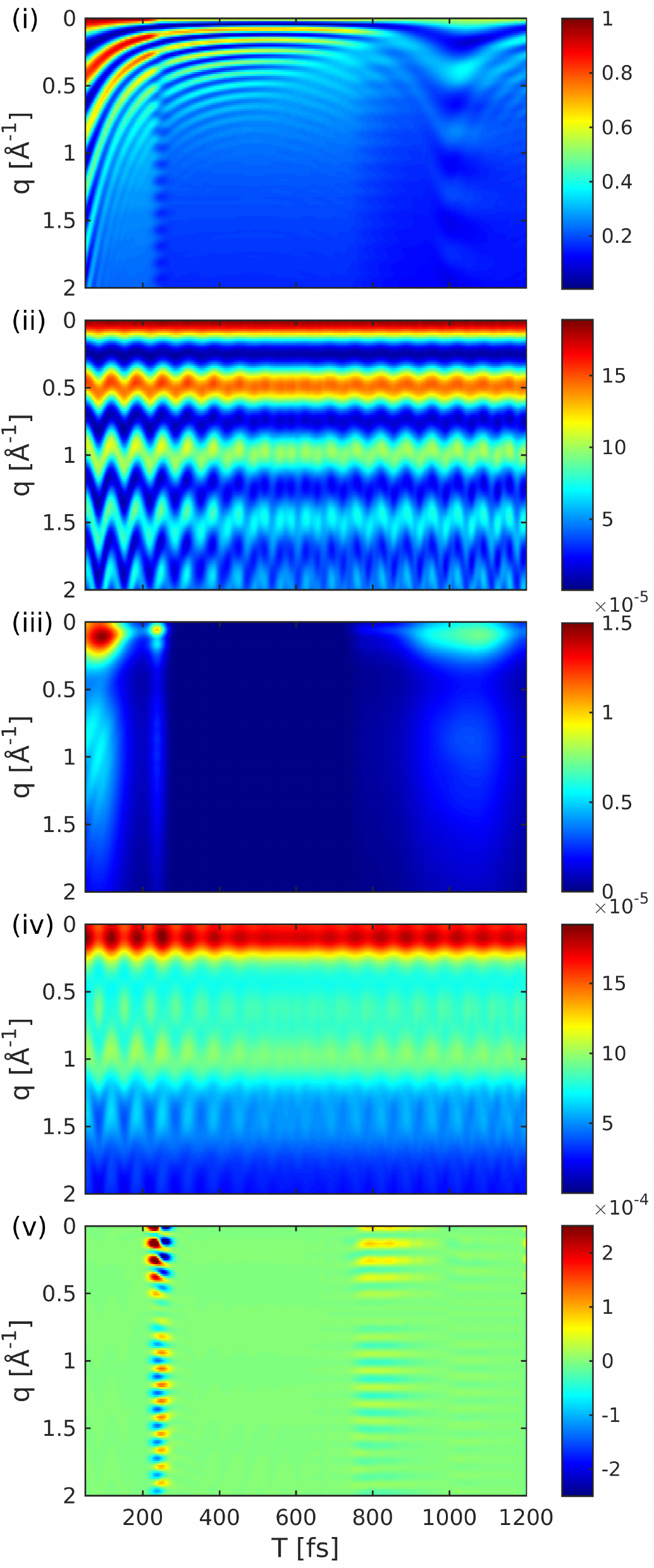}
\caption{Variation with probe delay $T$ of the five contributions to the gas-phase diffraction signal of NaF vs.\ the momentum transfer $q$. Panel labeling corresponds to  Eq.\ (\ref{eq:S1molExp}). Signal intensities are normalized
relative to (i).
(i) and (ii) elastic contributions from $e$ and $g$ respectively.
(iii) and (iv) weaker  inelastic contributions from $e$ and $g$ respectively.
(v) combined contribution of inelastic scattering and electronic coherences.
Its intensity lies between the elastic and inelastic terms.}\label{fig:S1}
\end{figure}

Figure\ \ref{fig:S1} shows the diffraction pattern as well as the relative magnitude of the five contributions to the signal in Eqs.\ (\ref{eq:S1molExp}) and (\ref{eq:Sinctr2}).
The contributions to the diffraction signal
are shown as labeled in Eq.\ (\ref{eq:S1molExp})
((i) through (v), corresponding to the diagrams of Fig.\ \ref{fig:Cdiag}).
The elastic diffraction signal, which stems from the charge density $\hat\sigma_{ee}$, is shown in Fig.\ \ref{fig:S1}(i). 
The time evolution represents the wave packet motion,
i.e., the fringe spacing increases as the wave packet moves towards a longer bond length.
The actinic pump-pulse (full width at half maximum 10\,fs) also creates a non-stationary nuclear wave packet in the electronic ground state.
This ground-state hole has comparable magnitude to the excited state contribution Fig.\ \ref{fig:S1}(i).
Figure \ref{fig:S1}(ii) shows the diffraction signal from the ground state density. The interference fringes are
signatures of an oscillating vibrational wave packet in the ground-state potential.
This hole burning phenomena will occur for pump-pulses that have bandwidths smaller than the
bandwidth of the Franck-Condon region.

The inelastic contribution that stems solely from the transition
densities $\hat\sigma_{eg}$ and the excited-state wavepacket in Fig.\ \ref{fig:S1}(iii), is  $\approx$ 4 orders of magnitude weaker.
It carries no information about the electronic coherence but is dominated by the shape and magnitude of the transition density $\hat\sigma_{eg}^2$ and is closely related to the transition dipole moment. This contribution varies widely over time since the nuclear wavepacket enters
a region, where the transition dipole vanishes.
The inelastic scattering from the ground state
shown in Fig.\ \ref{fig:S1}(iv), is also modulated by the wavepacket motion.
Compared to Fig.\ \ref{fig:S1}(iii), its intensity is only weakly modulated
since it never reaches a region where the transition dipole moment vanishes.

Figure \ref{fig:S1}(v) depicts the combined contribution of inelastic scattering of the electronic coherences. This contribution is responsible
for the time-evolving density caused by the electron dynamics
\cite{Dixit12pnas}.
At $\approx$ 220\,fs, when the wavepacket hits the avoided crossing regime,
an electronic coherence is created, resulting in a slow temporal oscillation that spreads over a wide range in $q$-space.
This contribution is $\approx$ 3 orders of magnitude weaker than the excited-state density (Fig.\ \ref{fig:S1}(i)) but one order of magnitude larger
than the other inelastic contributions (iii) and (iv).
Another contribution appears at around 800\,fs, which stems
from the returning wavepacket but it is much weaker due to the larger spread of the wavepacket. When the wavepacket returns to the Franck-Condon point, a larger
spike in the coherence is visible in Fig.\ \ref{fig:popcoh} at around 1100\,fs.
This contribution is averaged out in Fig.\ \ref{fig:S1}(v) due to the probe-pulse length and would require an attosecond rather than a femtosecond pulse to observe.
\begin{figure}
\centering
\includegraphics[width=0.5\textwidth]{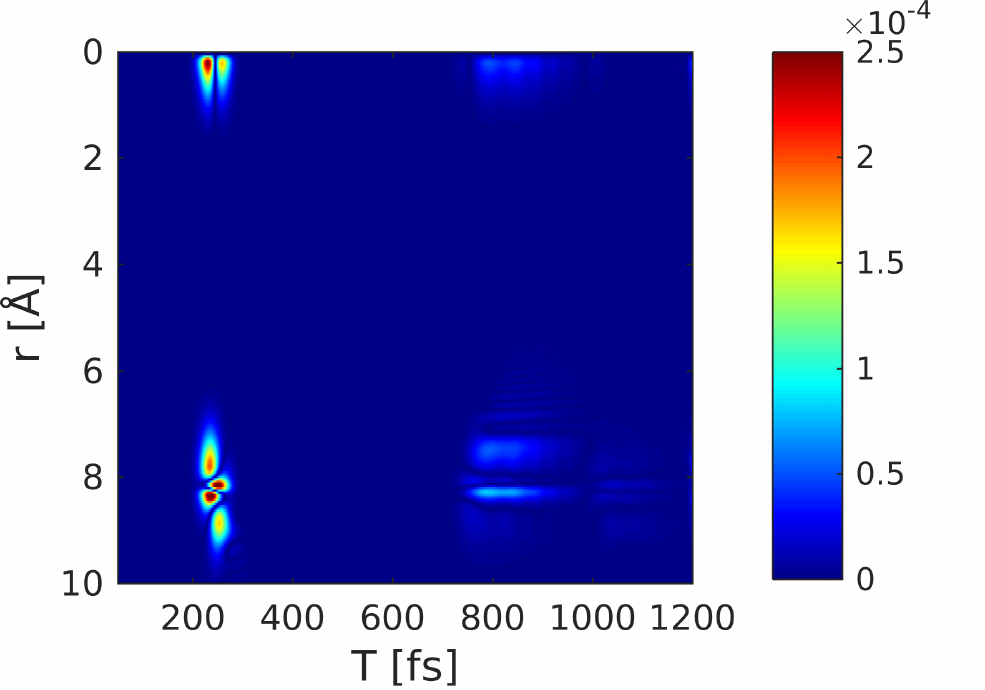}
\caption{Real-space picture of the coherence contribution
to the TRXD signal obtained by a Fourier transform of Fig.\ \ref{fig:S1}(v).}
\label{fig:S1inv}
\end{figure}
Figure \ref{fig:S1inv} shows the
coherence contribution in Fig.\ \ref{fig:S1}(v) in real-space (Fourier transform). The first passage
through the avoided crossing at 200\,fs shows a clear signature at 8\,\AA,
thus giving a hint of where the electronic coherence has been created.
The second passage at around 800\,fs is also visible.

\section{Conclusions}
In conclusion, the simulated gas-phase or single-molecule diffraction signal of sodium fluoride undergoing nonadiabatic avoided crossing dynamics contains
signatures of the created electronic coherence on top of the 
dominant ground- and excited-state wavepacket motion.
The diffraction signal depends on the
ground- and the excited-state charge densities as well as the
transition charge density that causes the inelastic contribution (v).
These densities depend on time through the interatomic distance, which
can be extracted directly from the diffraction signal.
The shape of the nuclear wavepacket can be qualitatively retrieved without further phase reconstruction.
For diatomic molecules, this allows to create a molecular movie out of the diffraction data.
The coherence contributions do not merely indicate that a coherence has been created but also reveal where it has been created. Its contribution is significantly weaker than elastic scattering processes and appears as a rapid oscillation on top of the diffraction pattern.
It will be interesting to explore other nonlinear optical signals where the coherence contribution is more pronounced and possibly background free \cite{Kowalewski15prl,Kowalewski16prl}. Finally, we note that, by including additional nuclear coordinates our approach may be used to predict signatures of CoIns in polyatomic molecules.
Extended nonlinear probe schemes may be capable of directly imaging the transition
density.

\acknowledgements 
The support of the Chemical Sciences, Geosciences, and Biosciences division, Office of Basic
Energy Sciences, Office of Science, U.S. Department of Energy through award
No.~DE-FG02-04ER15571 is gratefully acknowledged. Support for K.B. was provided by DOE. M.K. gratefully acknowledges support from the Alexander von Humboldt foundation through the Feodor Lynen program.

\appendix
\section{The Scattering Signal}\label{app:sig}
We start with the basic definition of the signal derived from time-dependent perturbation theory, which allows its convenient
dissection into its different contributions.
In a previous work\cite{bennett2014time}, we derived the following expressions for single-molecule frequency-resolved diffraction signals
\begin{widetext}
\begin{align}\label{eq:Sincwr}
S(\bar{\omega},\mathbf{k}_s,\Lambda)=\int d\omega\vert \mathcal{F}_f(\omega,\bar{\omega})\vert^2 \omega'^2\sum_{\alpha} &\int d\omega_pd\omega_{p'}A_p(\omega_p)A^*_p(\omega_{p'})e^{-i(\mathbf{q}-\mathbf{q}')\cdot\mathbf{r}_\alpha}  \\ & \notag\times \langle\hat{\sigma}_\alpha(-\mathbf{q}',\omega_{p'}-\omega')\hat{\sigma}_\alpha(\mathbf{q},\omega'-\omega_p)\rangle.
\end{align}
\end{widetext}
In Eq.\ (\ref{eq:Sincwr}), $A_p(\omega)$ is the vector potential envelope for the X-ray probe, $\mathcal{F}_f$ is a frequency gating (detector sensitivity) function, $\Lambda$ stands for the set of parameters defining the X-ray field, and $\mathbf{q}^{(\prime)}\equiv\frac{\omega}{c}\hat{\mathbf{k}_s}-\mathbf{k}_{p^{(\prime)}}$ is the momentum transfer ($\hat{\mathbf{k}_s}$ is the direction of the scattered wavevector).
In standard applications, the molecules that compose the sample are assumed to have identical charge distributions and the subscripts $\alpha$ on the charge density should be dropped, as we will do for the remainder of this manuscript. Assuming no frequency resolution, we have $\mathcal{F}_f(\omega,\bar{\omega})=1$ for the frequency gating function. The long-range (inter-molecular) structure of the sample is captured by the structure factors
\begin{align}
F_1(\mathbf{q})&=\sum_\alpha e^{-i\mathbf{q}\cdot\mathbf{r}_\alpha}
\end{align}
in terms of which the diffraction signals can be written as
\begin{align} \label{eq:Sincwr2}
S_{1}(\mathbf{k}_s,\Lambda)&=\int d\omega\frac{\omega^2}{\omega_p\omega_p'}\int d\omega_pd\omega_{p'}  E_p(\omega_p)E^*_p(\omega_{p'})\\ \notag &\times F_1(\mathbf{q}-\mathbf{q}')\langle\hat{\sigma}(-\mathbf{q}',\omega_{p'}-\omega)\hat{\sigma}(\mathbf{q},\omega-\omega_p)\rangle)\rangle \label{eq:Scohwr2}
\end{align}
where we have substituted the electric field envelopes for the vector potential.  For near-elastic scattering, we approximate $\frac{\omega}{\omega_p^{(\prime)}}\approx 1$, which is nearly valid even for inelasticities of several eV since the central frequency of the X-ray pulse is on the order of 10keV. Similarly, the momentum transfer is approximated as independent of frequency.  For the purposes of time-resolved diffraction studies, a time-domain expression is more convenient to simulate due to the nuclear motion. We thus substitute the time-domain charge density operator 
\begin{align}
\hat{\sigma}(\mathbf{q},\omega)=\int dt\hat{\sigma}(\mathbf{q},t)e^{i\omega t},
\end{align}
where we work in the interaction picture so that the operator time-dependence is through the field-free propagator, to obtain 
\begin{align}\label{eq:Scohtr}
S(\mathbf{q},\Lambda)&=F_1(0)\int d\omega\int dtdt'e^{i\omega(t-t')}  E_p(t)E^*_p(t')\langle\hat{\sigma}(-\mathbf{q},t')\hat{\sigma}(\mathbf{q},t)\rangle 
\end{align}
where we have replaced $\mathbf{k}_s$ by $\mathbf{q}$ in the argument in accordance with the quasi-elastic approximation. Upon carrying out the $d\omega$ integration and using $\hat{\sigma}(-\mathbf{q})=\hat{\sigma}^*(\mathbf{q})$, finally results in Eq.\ (\ref{eq:Sinctr2})

\section{The Electronic Charge Density Operator}
In this section, we discuss the operator nature of the charge density and its consequences.  In this section, we will ignore nuclear dependence and will begin by considering a one-electron system.  We seek an operator $\hat{\sigma}(\mathbf{r})$ such that the expectation value in a given state is the charge density
\begin{align}
\vert\psi(\mathbf{r})\vert^2&\equiv\langle\psi\vert\hat{\sigma}(\mathbf{r})\vert\psi\rangle=\int d\mathbf{r}'d\mathbf{r}''\langle\psi\vert\mathbf{r}''\rangle\langle\mathbf{r}''\vert\hat{\sigma}(\mathbf{r})\vert\mathbf{r}'\rangle\langle\mathbf{r}'\vert\psi\rangle\\ \notag &= \int d\mathbf{r}'d\mathbf{r}''\psi^*(\mathbf{r}'')\psi(\mathbf{r}')\langle\mathbf{r}''\vert\hat{\sigma}(\mathbf{r})\vert\mathbf{r}'\rangle
\end{align}
This identifies the real-space matrix elements of the electronic charge density field operator
\begin{align}\label{eq:1esigma}
\sigma_{\mathbf{r}''\mathbf{r}'}(\mathbf{r})\equiv\langle\mathbf{r}''\vert\hat{\sigma}(\mathbf{r})\vert\mathbf{r}'\rangle=\delta(\mathbf{r}-\mathbf{r}')\delta(\mathbf{r}-\mathbf{r}'').
\end{align}
For a state decomposed into eigenmodes $\vert\psi\rangle=\sum_k c_i\vert i\rangle$, we have
\begin{align}
\langle\psi\vert\hat{\sigma}(\mathbf{r})\vert\psi\rangle=\sum_{ij}\rho_{ij}\psi^*_i(\mathbf{r})\psi_j(\mathbf{r})
\end{align}
where $\rho_{ij}=c^*_ic_j$ are the electronic populations and coherences.  Note that this matches the usual field-theoretic definition of the charge density $\hat{\sigma}(\mathbf{r})=\hat{\psi}^\dagger(\mathbf{r})\hat{\psi}(\mathbf{r})$

\subsection{The One-Electron Charge Density Operator of a Many-Electron System}
In this section, we extend the reasoning of the previous section to an $n$-electron state $\vert \Psi\rangle$.  The real-space identity operator in the space spanned by such states is 
\begin{align}
\int d\mathbf{r}_1\dots d\mathbf{r}_n\vert\mathbf{r}_1,\dots,\mathbf{r}_n\rangle\langle\mathbf{r}_1,\dots,\mathbf{r}_n\vert \equiv\int \lbrace d\mathbf{r}\rbrace\big\vert\lbrace\mathbf{r}\rbrace\rangle\langle\lbrace\mathbf{r}\rbrace\big\vert
\end{align}
and the one-electron charge density is \cite{Szabo}
\begin{align}\label{eq:neExp}
&\int d\mathbf{r}_2\dots d\mathbf{r}_n\big\vert\Psi\left(\lbrace\mathbf{r}\rbrace\right)\big\vert^2=\langle\Psi\vert\hat{\sigma}(\mathbf{r})\vert\Psi\rangle\\ \notag=&\int \lbrace d\mathbf{r}'\rbrace\lbrace d\mathbf{r}''\rbrace\Psi^*(\lbrace \mathbf{r}''\rbrace)\Psi(\lbrace\mathbf{r}'\rbrace)\langle\lbrace\mathbf{r}''\vert\hat\sigma(\mathbf{r})\vert\lbrace\mathbf{r}'\rbrace\rangle
\end{align}
Since the charge-density operator is a one-electron operator, we have the straightforward $n$-electron generalization of Eq.\ (\ref{eq:1esigma})
\begin{align}
\langle\lbrace\mathbf{r}''\vert\hat\sigma(\mathbf{r})\vert\lbrace\mathbf{r}'\rbrace\rangle= \sum_m\delta(\mathbf{r}-\mathbf{r}'_l) \delta(\mathbf{r}-\mathbf{r}''_l)\prod_{m\ne l}\delta(\mathbf{r}'_m-\mathbf{r}''_m)
\end{align}
which is directly confirmed by substitution into Eq.\ (\ref{eq:neExp}) and gives
\begin{align}
\langle\Psi\vert\hat{\sigma}(\mathbf{r})\vert\Psi\rangle=\sum_{ij}\rho_{ij}\sigma_{ij}(\mathbf{r})
\end{align}
where we have identified
\begin{align}\label{eq:nesigma}
\sigma_{ij}(\mathbf{r})=\int d\mathbf{r}_2\dots d\mathbf{r}_n\Psi^*_i(\mathbf{r}_1,\dots\mathbf{r}_n)\Psi_j(\mathbf{r}_1,\dots\mathbf{r}_n)
\end{align}
We note that this result can equally well be obtained by use of real-space field operators for many-electron systems as explicated by Cederbaum \cite{cederbaum,Cederbaum16jpca}.  Moreover, Eq.\ (\ref{eq:nesigma}) is readily generalized to account for nuclear degrees of freedom $\mathbf{R}$ as 
\begin{align}
\hat{\sigma}_{ij}(\mathbf{r})=\int d\mathbf{r}_2\dots d\mathbf{r}_n\Psi^*_i(\mathbf{R},\mathbf{r}_1,\dots\mathbf{r}_n)\Psi_j(\mathbf{R},\mathbf{r}_1,\dots\mathbf{r}_n)
\end{align}
where the circumflex indicates that the left hand side remains an operator in the nuclear subspace due to dependence on $\mathbf{R}$

\bibliography{TRXS,XRD}
\end{document}